# Tokamak elongation – how much is too much? I Theory

## J. P. Freidberg[1], A. Cerfon[2], J. P. Lee[1,2]

### Abstract

In this and the accompanying paper the problem of the maximally achievable elongation $\kappa$ in a tokamak is investigated. The work represents an extension of many earlier studies, which were often focused on determining $\kappa$ limits due to (1) natural elongation in a simple applied pure vertical field or (2) axisymmetric stability in the presence of a perfectly conducting wall. The extension investigated here includes the effect of the vertical stability feedback system which actually sets the maximum practical elongation limit in a real experiment. A basic resistive wall stability parameter $\gamma \tau_w$ is introduced to model the feedback system which although simple in appearance actually captures the essence of the feedback system. Elongation limits in the presence of feedback are then determined by calculating the maximum $\kappa$ against $n = 0$ resistive wall modes for fixed $\gamma \tau_w$. The results are obtained by means of a general formulation culminating in a variational principle which is particularly amenable to numerical analysis. The principle is valid for arbitrary profiles but simplifies significantly for the Solov'ev profiles, effectively reducing the 2-D stability problem into a 1-D problem. The accompanying paper provides the numerical results and leads to a sharp answer of "how much elongation is too much"?

1. Plasma Science and Fusion Center, MIT, Cambridge MA
2. Courant Institute of Mathematical Sciences, NYU, New York City NY

## 1. Introduction

It has been known for many years that tokamak performance, as measured by pressure and energy confinement time, improves substantially as the plasma cross section becomes more elongated. There are, however, also well known limits on the maximum achievable elongation, which arise from the excitation of $n = 0$ vertical instabilities. When designing next generation reactor scale tokamak experiments [Aymar et al. 2002; Najmabadi et al. 1997; Najmabadi et al. 2006; Sorbom et al. 2015], where high performance is critical, it is thus important to be able to accurately predict the maximum achievable elongation $\kappa$ as a function of inverse aspect ratio $\varepsilon = a \, / \, R_0$, where $a$ is the minor radius of the device, and $R_0$ is the major radius.

The inverse aspect ratio is a particularly important parameter since different reactor designs have substantially different values for this quantity. Specifically, standard and high field tokamak reactor designs have $1 \, / \, \varepsilon \sim 3 - 4.5$ [Aymar et al. 2002; Najmabadi et al. 1997; Najmabadi et al. 2006; Sorbom et al. 2015; Schissel et al. 1991; Hutchinson et al. 1994], while spherical tokamaks have smaller $1 \, / \, \varepsilon \sim 1.5$ [Sabbagh et al. 2001; Sykes et al. 2001]. Since optimized plasma performance and corresponding minimized cost depend strongly on $\kappa$ it is important to have an accurate determination of $\kappa = \kappa(\varepsilon)$ and this is the goal of the present and accompanying paper.

The problem of determining $\kappa = \kappa(\varepsilon)$ has received considerable attention in past studies but, as discussed below, there is still an important gap in our knowledge. To put the problem in perspective we note that earlier studies generally fall into one of three main categories, each one providing valuable information, but not the whole story. These are summarized as follows.

The first class of studies involves the concept of "natural elongation". Here, the tokamak is immersed in a deliberately simple external poloidal magnetic field, usually a pure vertical field. The plasma is expected to be stable against $n = 0$ vertical instabilities since one vertical position is the same as any other. In these studies [Peng and Strickler 1986; Roberto and Galvão 1992] the free boundary equilibrium surface is calculated (i.e. the elongation $\kappa$ and the triangularity $\delta$ are determined) for a range of $\varepsilon$ from which it is straightforward to extract the desired $\kappa = \kappa(\varepsilon)$. Excellent physical insight is provided by these calculations. Even so, the maximum predicted natural elongations are substantially smaller than those achieved in experiments. The reason is that experiments have feedback systems that can stabilize $n = 0$ vertical instabilities, thereby allowing larger $\kappa$'s than those predicted by marginally stable naturally elongated configurations.

The second class of studies involves the calculation of the critical normalized wall radius $b \, / \, a$ for an ideal perfectly conducting conformal wall required to stabilize a desired elongation; that is the analysis basically determines $b \, / \, a = f(\varepsilon, \kappa)$. Here too, the studies [Wesson 1978] (and references therein) provide valuable insight. Specifically, the studies determine the maximum $b/a$ that might potentially be able to be stabilized by feedback. However, the analysis does not actually predict whether or not a practical feedback system can be constructed to provide stability. Equally important, from a practical point of view the actual value of $b/a$ does not vary much from experiment to experiment.

The third class of studies consists of detailed, engineering level designs that predict the maximum elongation and which include many effects such as plasma profiles, real geometry, safety margins, and most importantly, engineering properties of the feedback system. Such studies [Kessel et al. 2006] are very realistic and are exactly what is required to design an actual experiment. Nevertheless, these are point designs whose main goals are focused on a specific machine rather than providing scaling insight in the form of $\kappa = \kappa(\varepsilon)$.

The present analysis attempts to fill an important gap in our knowledge, namely determining the maximum elongation as a function of inverse aspect ratio including the constraints arising from the feedback system. The calculation is largely analytic combined with some straightforward numerics. In our model the plasma is assumed to be up-down symmetric and is characterized by the following parameters: inverse aspect ratio $\varepsilon = a/R_0$, elongation $\kappa$, triangularity $\delta$, poloidal beta $\beta_p$, normalized wall radius $b/a$, and an appropriately defined parameter representing the capabilities of the feedback system. Defining this feedback parameter and including it in the analysis is the main new contribution of the research.

A simple but reliable definition of the feedback parameter is based on the following two observations. First, practical feedback is feasible when the growth rate $\gamma$ of the $n=0$ vertical instability is small. Since the plasma is surrounded by a finite conductivity wall the instability of interest is actually a resistive wall mode. The second observation is that if the feedback coils are located outside the resistive wall, as they usually are, then an effective feedback system must have a rapid field diffusion time $\tau_w$. This is important because once an unstable plasma motion is detected the feedback response fields must quickly diffuse through the wall in order to reach the plasma. In other words $\tau_w$ should be short.

In principle it is possible to trade off growth rate in favor of response time or vice-versa. However, the overall effectiveness of the feedback system is dependent upon the combined smallness of $\gamma$ and $\tau_w$. The conclusion is that a simple parameter that takes into account the feedback system is the product of these two quantities:

$$\gamma \tau_w \quad = \quad \text{feedback capability parameter} \qquad (1)$$

A good way to think about this parameter is as follows. In designing a new experiment the properties of the feedback system depend mainly on the geometry of the vacuum chamber, the availability of fast-response power supplies, the maximum feedback power available, the number of feedback coils, and the sensitivity of the detectors. These represent a combination of engineering and economic constraints that should not vary much as $\varepsilon$ changes. Consequently, the engineering value of $\gamma \tau_w$ is a good measure of the feedback capability. It represents the maximum value of the resistive wall growth rate that can be feedback stabilized.

Now, consider the design of a new experiment with $\varepsilon$ as a parameter over which optimization is to be performed. For a fixed $\varepsilon$ the plasma elongation should be increased until the growth rate of the $n = 0$ resistive wall mode is equal to the value of the feedback capability parameter given in Eq. (1). In this process, the triangularity $\delta$ may also be adjusted to find the maximum allowable $\kappa$. This resulting $\kappa$ is the elongation that maximizes performance for the given $\varepsilon$. By repeating the procedure for different $\varepsilon$ it is then possible to determine the optimum aspect ratio, including the effects of transport, heating, magnetic design, etc., that results in the peak value of maximum performance and corresponding minimizes cost. In our analysis practical engineering values of $\gamma \tau_w$ are chosen by examining the data from existing experiments as well as ITER [Aymar et al. 2002].

Based on this discussion we can state that the goals of the analysis are to determine the maximum allowable elongation $\kappa$ and corresponding triangularity $\delta$ as a function of inverse aspect ratio $\varepsilon$ subject to the constraints of fixed poloidal beta $\beta_p$, normalized wall radius $b / a$, and feedback parameter $\gamma \tau_w$; that is we want to determine

$$\kappa = \kappa(\varepsilon; \beta_p, b / a, \gamma \tau_w)$$
$$\delta = \delta(\varepsilon; \beta_p, b / a, \gamma \tau_w)$$

(2)

This paper presents the analytic theory necessary to determine these quantities. The analysis is valid for arbitrary plasma profiles. An additional useful result presented as the derivation progresses is an explicit relationship between vertical stability and neighboring equilibria of the Grad-Shafranov equation, a relationship long believed to be true but to the authors' knowledge, never explicitly appearing in the literature.

The second paper in the two part series presents the numerical results. For numerical simplicity the results are obtained for the Solov'ev profiles [Solov'ev 1968] although with some additional equilibrium numerical work it is possible to include arbitrary profiles. One may expect Solov'ev profiles to provide reliable physical insight since axisymmetric MHD modes are thought to not depend too sensitively on the details of the current profile [Bernard et al. 1978]. The overall conclusions are that the maximum elongation (1) decreases substantially as $\varepsilon$ becomes smaller and (2) is substantially higher than that predicted by natural elongation calculations, much closer to what is observed experimentally. The maximum elongation is weakly dependent on $\beta_p$ and also does not vary much with $b/a$ primarily because this quantity itself has only limited variation in practical designs. There is, however, a substantial increase in maximum elongation as $\gamma\tau_w$ increases. As expected, feedback is very effective in increasing elongation and overall plasma performance. A slightly more subtle effect is the value of the optimum triangularity corresponding to maximum elongation, which is noticeably smaller than that observed experimentally in high performance plasma discharges. The reason is presumably associated with the fact that maximum overall performance depends on turbulent energy transport as well as MHD stability. Although turbulent transport is known to be reduced with increasing triangularity, which helps explain the data, it is unfortunate that the current empirical scaling laws do not explicitly include this effect.

The analysis is now ready to proceed.

## 2. Equilibrium

The equilibrium of an axisymmetric tokamak is described by the well-known Grad-Shafranov equation [Grad and Rubin 1958; Shafranov 1958]

$$\Delta^{*}\Psi = -\mu_0 R^2 \frac{dp}{d\Psi} - \frac{1}{2}\frac{dF^2}{d\Psi} \tag{3}$$

Here, $\Psi(R, Z) =$ poloidal flux$/2\pi$ and $p(\Psi)$, $F(\Psi)$ are two free functions. The $\Delta^{*}$ operator, plus the magnetic field $\mathbf{B} = \mathbf{B}_p + B_\phi \mathbf{e}_\phi$ and current density $\mathbf{J} = \mathbf{J}_p + J_\phi \mathbf{e}_\phi$ are given by

$$\Delta^{*}\Psi \equiv R\frac{\partial}{\partial R}\left(\frac{1}{R}\frac{\partial\Psi}{\partial R}\right) + \frac{\partial^2\Psi}{\partial Z^2}$$

$$\mathbf{B} = \frac{1}{R}\nabla\Psi\times\mathbf{e}_\phi + \frac{F}{R}\mathbf{e}_\phi \tag{4}$$

$$\mu_0\mathbf{J} = \frac{dF}{d\Psi}\mathbf{B}_p - \frac{1}{R}\Delta^{*}\Psi\,\mathbf{e}_\phi$$

For the study of $n = 0$ resistive wall modes the essential physics is captured by considering up-down symmetric equilibria and this is the strategy adopted here. Also, great analytic simplicity follows by choosing the free functions to correspond to the Solov'ev profiles [Solov'ev 1968] for which $FF'(\Psi) = C_1$ and $p'(\Psi) = C_2$, with prime denoting $d/d\Psi$. Only two constants, $C_1$ and $C_2$, are needed to specify an equilibrium.

The Grad-Shafranov equation can now be conveniently normalized in terms of two equivalent constants, $\Psi_0$ and $A$, one of which ($\Psi_0$) scales out entirely from the final formulation. The full set of normalizations is given by

$$R = R_0 X$$

$$Z = R_0 Y$$

$$\Psi = \Psi_0 \mathit{\Psi} \qquad (5)$$

$$\mu_0 \frac{dp}{d\Psi} = -\frac{\Psi_0(1-A)}{R_0^4}$$

$$F \frac{dF}{d\Psi} = -\frac{\Psi_0 A}{R_0^2}$$

Note that the normalized flux has an upper case italic font. In terms of these normalizations the Grad-Shafranov equation plus the critical field quantities needed for the analysis reduce to

$$\mathit{\Psi}_{XX} - \frac{1}{X}\mathit{\Psi}_X + \mathit{\Psi}_{YY} = A + (1-A)X^2$$

$$(R_0^2 / \Psi_0)\mathbf{B}_p = \frac{1}{X}\left(-\mathit{\Psi}_Y \mathbf{e}_R + \mathit{\Psi}_X \mathbf{e}_Z\right) \qquad (6)$$

$$(R_0^3 / \Psi_0)\mu_0 J_\phi = -\frac{1}{X}\left[A + (1-A)X^2\right]$$

The boundary conditions require regularity in the plasma and $\mathit{\Psi}(S_P) = 0$ with $S_P$ the plasma boundary. This implies that $\mathit{\Psi}(X,Y) < 0$ in the plasma volume. Typical values of the free constant $A$ in decreasing order are as follows

$$A = 1 \qquad\qquad p(\Psi) = 0 \text{ (force free)}$$

$$A = 0 \qquad\qquad F(\Psi) = R_0 B_0 \text{ (vacuum toroidal field)} \qquad (7)$$

$$A = -\frac{(1-\varepsilon)^2}{1-(1-\varepsilon)^2} \qquad J_\phi(1-\varepsilon, 0) = 0 \text{ (inner edge current reversal)}$$

The task now is to find a solution for $\mathit{\Psi}(X,Y)$. Exact solutions to the Grad-Shafranov equation for Solov'ev profiles have been derived by a number of authors [Zheng et al. 1996; Weening 2000; Shi 2005]. Here, we follow the formulation of reference [Cerfon and Freidberg 2010; Freidberg 2014]. Specifically, an exact solution to Eq. (6) can be written as

$$\Psi(X,Y) = \frac{X^4}{8} + A\left(\frac{1}{2}X^2 \ln X - \frac{X^4}{8}\right) + c_1\Psi_1 + c_2\Psi_2 + c\Psi_3 + c_4\Psi_4 + c_5\Psi_5 + c_6\Psi_6 + c_7\Psi_7$$

$$\Psi_1 = 1$$

$$\Psi_2 = X^2$$

$$\Psi_3 = Y^2 - X^2 \ln X \qquad\qquad (8)$$

$$\Psi_4 = X^4 - 4X^2Y^2$$

$$\Psi_5 = 2Y^4 - 9X^2Y^2 + 3X^4 \ln X - 12X^2Y^2 \ln X$$

$$\Psi_6 = X^6 - 12X^4Y^2 + 8X^2Y^4$$

$$\Psi_7 = 8Y^6 - 140X^2Y^4 + 75X^4Y^2 - 15X^6 \ln X + 180X^4Y^2 \ln X - 120X^2Y^4 \ln X$$

The free constants $c_j$ are chosen to match as closely as possible the well-known analytic model for a smooth, elongated, "D" shaped boundary cross section $S_p$ given parametrically in terms of $\tau$ by [Miller et al. 1998]

$$X = 1 + \varepsilon \cos(\tau + \delta_0 \sin \tau)$$

$$Y = \varepsilon \kappa \sin \tau \qquad\qquad (9)$$

Here, $\varepsilon = a/R_0$ is the inverse aspect ratio, $\kappa$ is the elongation, and $\delta = \sin \delta_0$ is the triangularity. The geometry is illustrated in Fig. 1.

In practice the $c_j$ are determined by requiring that the exact plasma surface $\Psi = 0$, plus its slope and curvature match the model surface at three points: (1) the outer equatorial point, (2) the inner equatorial point, and (3) the high point maximum. Taking into account the up-down symmetry we see that the requirements translate to

$$\Psi(1+\varepsilon,0) = 0 \qquad\qquad \text{outer equatorial point}$$

$$\Psi(1-\varepsilon,0) = 0 \qquad\qquad \text{inner equatorial point}$$

$$\Psi(1-\delta\varepsilon,\kappa\varepsilon) = 0 \qquad\qquad \text{high point}$$

$$\Psi_X(1-\delta\varepsilon,\kappa\varepsilon) = 0 \qquad\qquad \text{high point maximum} \qquad (10)$$

$$\Psi_{YY}(1+\varepsilon,0) = -N_1\Psi_X(1+\varepsilon,0) \qquad\qquad \text{outer equatorial point curvature}$$

$$\Psi_{YY}(1-\varepsilon,0) = -N_2\Psi_X(1-\varepsilon,0) \qquad\qquad \text{inner equatorial point curvature}$$

$$\Psi_{XX}(1-\delta\varepsilon,\kappa\varepsilon) = -N_3\Psi_Y(1-\delta\varepsilon,\kappa\varepsilon) \qquad\qquad \text{high point curvature}$$

where for the given model surface,

$$N_1 = \left[\frac{d^2X}{dY^2}\right]_{\tau=0} = -\frac{\left(1+\delta_0\right)^2}{\varepsilon\kappa^2}$$

$$N_2 = \left[\frac{d^2X}{dY^2}\right]_{\tau=\pi} = \frac{\left(1-\delta_0\right)^2}{\varepsilon\kappa^2} \qquad (11)$$

$$N_3 = \left[\frac{d^2Y}{dX^2}\right]_{\tau=\pi/2} = -\frac{\kappa}{\varepsilon\cos^2\delta_0}$$

The determination of the $c_j$ has been reduced to finding the solution to a set of seven linear, inhomogeneous algebraic equations, a very simple numerical problem. Hereafter, we assume that the $c_j$ have been determined, thereby completely defining the equilibrium solution. A typical set of flux surfaces corresponding to the high $\beta$, tight aspect ratio NSTX spherical tokamak is illustrated in Fig. 2, where $A$ has been chosen so that $J_\phi(1-\varepsilon,0) = 0$. The surfaces look quite reasonable and as expected, exhibit a large shift of the magnetic axis.

The one final piece of information required for the resistive wall stability analysis is the value of poloidal beta $\beta_p$. In general $\beta_p$ and the toroidal current $I$ can be expressed in terms of the free constants $A, \Psi_0$. The advantage of the normalizations introduced by Eq. (5) is that $\beta_p$ is only a function of $A$ but not $\Psi_0$. This can be seen by noting that the average pressure and toroidal current can be written as,

- Average pressure:

$$\overline{p} \equiv \frac{\int p \, d\mathbf{r}}{\int d\mathbf{r}} = -\frac{2\pi \Psi_0^2 (1-A)}{\mu_0 R_0 V} \int \Psi X \, dX \, dY$$

$$V = \int d\mathbf{r} = 2\pi R_0^3 \int X \, dX \, dY \tag{12}$$

- Toroidal current:

$$\mu_0 I = \oint_{\Psi=0} B_p \, dl = \int_{\Psi \leq 0} \mu_0 J_\phi \, dS = -\frac{\Psi_0}{R_0} \int_{\Psi \leq 0} \frac{1}{X} \big[ A + (1-A)X^2 \big] dX dY \tag{13}$$

where $B_p$ is the magnitude of the poloidal magnetic field. We now define $\beta_p$ as

$$\beta_p \equiv \frac{2\mu_0 \overline{p}}{\overline{B}_p^2}$$

$$\overline{B}_p = \frac{\oint\limits_{\Psi=0} B_p \, dl}{\oint\limits_{\Psi=0} dl} = \frac{\mu_0 I}{L_P}$$

$$L_P = \oint_{\Psi=0} dl \tag{14}$$

Substituting for $\overline{p}$ and $I$ yields

$$\beta_p = -\frac{4\pi R_0 L_P^2}{V} \frac{(1-A) \int\limits_{\Psi \leq 0} \Psi X \, dX \, dY}{\left\{ \int\limits_{\Psi \leq 0} \frac{1}{X} \big[ A + (1-A)X^2 \big] dX \, dY \right\}^2} \tag{15}$$

Since $\Psi$ is a function of $A$ but not $\Psi_0$ it follows that as stated $\beta_p = \beta_p(A, \varepsilon, \kappa, \delta)$. Note also that the volume $V$ and poloidal plasma circumference $L_P$ are purely geometric factors that can be calculated from the solution given by Eq. (8) once the $c_j$ have been determined. Furthermore, with our normalizations the ratio $R_0 L_P^2 / V$ is independent of $R_0$ and $\Psi_0$; that is $R_0 L_P^2 / V = f(A, \varepsilon, \kappa, \delta)$. This completes the discussion of the equilibrium.

## 3. Resistive wall stability formulation

In this section we derive a variational principle describing the stability of $n = 0$ resistive wall modes in a tokamak. The analysis starts from the ideal MHD normal mode equation for plasma stability. To obtain the variational principle we will need to decompose the volume surrounding the plasma into three regions: an interior vacuum region, a thin wall, and an exterior vacuum region [Haney and Freidberg 1989]. The final variational principle is expressed in terms of three perturbed flux functions for the (1) plasma, (2) interior vacuum region and (3) exterior vacuum region. To obtain this variational principle use is made of two natural boundary conditions.

The final step in the formulation is to derive relationships between the two perturbed vacuum fluxes and the perturbed plasma flux, a task accomplished by the application of Green's theorem. The analysis is somewhat simplified by focusing on up-down symmetric equilibria and considering only vertical-type displacements which are the most dangerous experimentally. The final form of the variational principle is straightforward to evaluate numerically and most importantly directly takes into account the feedback constraint $\gamma \tau_w =$ constant as described in the Introduction.

The variational formulation, when implemented numerically, allows us to determine the maximum elongation and corresponding triangularity as a function of aspect ratio.

- **The basic ideal MHD stability equations**

The starting point for the analysis is the ideal MHD linear stability equations for the plasma given by

$$\omega^2 \rho \boldsymbol{\xi} + \mathbf{F}(\boldsymbol{\xi}) = 0$$

$$\mathbf{F}(\boldsymbol{\xi}) = \mathbf{J}_1 \times \mathbf{B} + \mathbf{J} \times \mathbf{B}_1 - \nabla p_1$$

$$\mathbf{B}_1 = \nabla \times (\boldsymbol{\xi}_\perp \times \mathbf{B}) \qquad (16)$$

$$\mu_0 \mathbf{J}_1 = \nabla \times \nabla \times (\boldsymbol{\xi}_\perp \times \mathbf{B})$$

$$p_1 = -\boldsymbol{\xi}_\perp \cdot \nabla p - \gamma p \nabla \cdot \boldsymbol{\xi}$$

where $\boldsymbol{\xi}$ is the plasma displacement vector and quantities with a 1 subscript represent first order perturbations [Freidberg 2014]. For resistive wall modes the inertial effects can be neglected because the corresponding growth rates are very slow compared to the characteristic MHD time $a / V_{Ti}$ where $V_{Ti}$ is the ion thermal speed. Thus, the plasma behavior is described by

$$\mathbf{F}(\boldsymbol{\xi}) = 0 \qquad (17)$$

In other words, referring to the general ideal MHD formulation of the energy principle [Freidberg 2014], only $\delta W(\boldsymbol{\xi}, \boldsymbol{\xi})$ is needed to describe the plasma behavior, and $\omega^2 K(\boldsymbol{\xi}, \boldsymbol{\xi})$ can be ignored.

Next, form the total energy integral $\delta W$ for $n = 0$ modes in the usual way:

$$\delta W = -\frac{1}{2} \int \boldsymbol{\xi} \cdot \mathbf{F}(\boldsymbol{\xi}) d\mathbf{r} = 0 \tag{18}$$

Using standard analysis [21] we can rewrite Eq. (18) as follows

$$\delta W(\boldsymbol{\xi}, \boldsymbol{\xi}) = \delta W_F(\boldsymbol{\xi}, \boldsymbol{\xi}) + BT = 0$$

$$\delta W_F(\boldsymbol{\xi}, \boldsymbol{\xi}) = \frac{1}{2\mu_0} \int [(\mathbf{Q})^2 + \gamma \mu_0 p (\nabla \cdot \boldsymbol{\xi})^2 - \mu_0 \boldsymbol{\xi}_\perp \cdot \mathbf{J} \times \mathbf{Q} + \mu_0 (\boldsymbol{\xi}_\perp \cdot \nabla p)(\nabla \cdot \boldsymbol{\xi}_\perp)] d\mathbf{r} \tag{19}$$

$$BT = \frac{1}{2\mu_0} \int_{S_P} (\mathbf{n} \cdot \boldsymbol{\xi}_\perp)[\mathbf{B} \cdot \mathbf{Q} - \gamma \mu_0 p \nabla \cdot \boldsymbol{\xi} - \mu_0 \boldsymbol{\xi}_\perp \cdot \nabla p] dS_P$$

where $\mathbf{Q} = \nabla \times (\boldsymbol{\xi}_\perp \times \mathbf{B})$, $\mathbf{n}$ is the outward unit surface normal vector, and the surface integral in $BT$ is evaluated on the plasma surface $S_P$ given by $\Psi = 0$. This form is general, and in particular is valid for $n = 0$ modes in a tokamak with arbitrary profiles.

Many researchers have long believed that $n = 0$ stability is closely related to the problem of neighboring equilibria of the Grad-Shafranov equation. This is indeed a correct belief although to the authors' knowledge a derivation of this connection has not appeared in the literature. We have derived a relationship which explicitly shows this connection. The analysis is somewhat lengthy and is given in Appendix A. The final result is a simplified form of $\delta W_F$ valid for up-down symmetric tokamaks undergoing vertical-like displacements; that is, displacements for which $\xi_Z(R, Z) = \xi_Z(R, -Z)$ and $\xi_R(R, Z) = -\xi_R(R, -Z)$, (e.g. $\xi_Z = \xi_0 =$ constant, $\xi_R = 0$). Because of up-down symmetry these displacements exactly decouple from the horizontal-like displacements for which $\xi_Z(R, Z) = -\xi_Z(R, Z), \xi_R(R, Z) = \xi_R(R, -Z)$. In addition, it is shown in Appendix A that the most unstable displacements are incompressible: $\nabla \cdot \boldsymbol{\xi} = 0$.

The simplified form of $\delta W_F$ is conveniently expressed in terms of the perturbed plasma flux $\psi$ which is related to the displacement vector by

$$\psi = -\boldsymbol{\xi}_\perp \cdot \nabla \Psi \tag{20}$$

The resulting form of $\delta W_F$, valid for arbitrary equilibrium profiles, is given by

$$\delta W_F = \frac{1}{2\mu_0} \int \left[ \frac{(\nabla \psi)^2}{R^2} - \left( \mu_0 p'' + \frac{1}{2R^2} F^{2''} \right) \psi^2 \right] d\mathbf{r} + \frac{1}{2} \int_{S_P} \left( \frac{\mu_0 J_\phi}{R^2 B_p} \psi^2 \right) dS_P \qquad (21)$$

Where, as before, a prime denotes a derivative with respect to the poloidal flux function, i.e. $d / d\Psi$. Note that $\delta W_F$ is already written in self adjoint form.

The connection to Grad-Shafranov neighboring equilibria is now apparent. The differential equation in $\psi$ obtained by setting the variation in $\delta W_F$ to zero is identical to the neighboring equilibrium equation obtained by letting $\Psi \to \Psi + \psi$ in the Grad-Shafranov equation and setting the first order contribution to zero.

Observe that there is a boundary term arising from several integrations by parts in the derivation. This term is often zero since $J_\phi(S) = 0$ on the plasma surface for many realistic profiles. However, it is not zero for the Solov'ev profiles since the edge current density is finite and is, in fact, is the main drive for vertical instabilities. To compensate this difficulty note that $p'' = F^{2''} = 0$ for the Solov'ev profiles implying that the volume contribution to $\delta W_F$ reduces to that of a vacuum field. In other words, the perturbed toroidal current density in the plasma is zero for the Solov'ev profiles. This simplification greatly outweighs the difficulty of maintaining the surface contribution. Specifically, it will ultimately allow us to make use of an analytic form of the Green's function for the plasma region when evaluating $dW_F$, a mathematical advantage that does not apply to general profiles.

The task now is to evaluate and simplify the boundary term $BT$, recasting it in a form that is automatically self-adjoint.

- **The inner vacuum region**

The first step in the simplification of $BT$ in Eq. (19) focuses on the inner vacuum region between the plasma and the resistive wall. By making use of incompressibility and rewriting Eq. (20) as $\mathbf{n} \cdot \boldsymbol{\xi}_\perp = \psi / RB_p$ where $\mathbf{n}$ is the unit normal vector to the plasma surface, we see that the boundary term reduces to

$$BT = -\frac{1}{2\mu_0} \int \frac{\psi}{RB_p} (\mathbf{B} \cdot \mathbf{Q} - \mu_0 \boldsymbol{\xi}_\perp \cdot \nabla p) dS_P \qquad (22)$$

Next, assume that the perturbed magnetic field in the inner vacuum region is also written in terms of a flux function $\hat{\psi}_i$. This is convenient because the previous analysis for the plasma region in terms of $\psi$ is directly applicable to the inner vacuum region by simply setting the equilibrium $\mathbf{J} = p = 0$ in the vacuum region.

Now, the jump conditions across $S_P$ require that (for incompressible displacements and no surface currents) [21]

$$
\begin{aligned}
\left.\hat{\psi}_i\right|_{S_P} &= \left.\psi\right|_{S_P} \\
\left.\hat{\mathbf{B}} \cdot \hat{\mathbf{B}}_i\right|_{S_P} &= \left.(\mathbf{B} \cdot \mathbf{Q} - \mu_0 \boldsymbol{\xi}_\perp \cdot \nabla p)\right|_{S_P}
\end{aligned}
\tag{23}
$$

with $\hat{\mathbf{B}}$ the equilibrium vacuum field in the inner vacuum region. At this point the first natural boundary condition is introduced into the formulation by substituting Eq. (23) into Eq. (22)

$$
BT = -\frac{1}{2\mu_0} \int_{S_P} \frac{\hat{\psi}_i}{RB_p} (\hat{\mathbf{B}} \cdot \hat{\mathbf{B}}_i) dS
\tag{24}
$$

Continuing, in the inner vacuum region the total (i.e. equilibrium plus perturbation) magnetic field can be written as

$$
\hat{\mathbf{B}}_{Tot} = \frac{1}{R} \nabla(\hat{\Psi} + \hat{\psi}_i) \times \mathbf{e}_\phi + \frac{\hat{F}(\hat{\Psi} + \hat{\psi}_i)}{R} \mathbf{e}_\phi
\tag{25}
$$

Since $\hat{F}(\hat{\Psi} + \hat{\psi}_i) = R_0 B_0 = $ constant in a vacuum region it follows that the perturbation $\hat{F}_i = (\partial \hat{F} / \partial \hat{\Psi})\hat{\psi}_i = 0$. Thus the perturbed magnetic field in the inner vacuum region for an $n = 0$ perturbation is given by

$$
\hat{\mathbf{B}}_i = \frac{1}{R} \nabla \hat{\psi}_i \times \mathbf{e}_\phi
\tag{26}
$$

from which it follows that

$$
\left.\hat{\mathbf{B}} \cdot \hat{\mathbf{B}}_i\right|_{S_P} = \left.\frac{1}{R^2} (\nabla \Psi \cdot \nabla \hat{\psi}_i)\right|_{S_P} = \left.\frac{B_p}{R} (\mathbf{n} \cdot \nabla \hat{\psi}_i)\right|_{S_P}
\tag{27}
$$

Here we have used the equilibrium continuity relation $\hat{\mathbf{B}}(S_P) = \mathbf{B}(S_P)$ which is valid when no surface currents are present. The boundary term can thus be written as

$$BT = -\frac{1}{2\mu_0} \int_{S_P} \frac{1}{R^2} \hat{\psi}_i (\mathbf{n} \cdot \nabla \hat{\psi}_i) dS_P \qquad (28)$$

The last part of this first step is to recognize that the magnetic energy in the inner vacuum region (with subscript $i$) can be written as

$$\delta W_{V_i} = \frac{1}{2\mu_0} \int (\hat{\mathbf{B}}_i)^2 \, d\mathbf{r} = \frac{1}{2\mu_0} \int \frac{(\nabla \hat{\psi}_i)^2}{R^2} \, d\mathbf{r} \qquad (29)$$

with $\hat{\psi}_i$ satisfying

$$\Delta^* \hat{\psi}_i = 0 \qquad (30)$$

Using Eq. (30) we can easily convert Eq. (29) into two surface integrals, one on the plasma surface $S_P$ and the other on the inner wall surface $S_W$,

$$\delta W_{V_i} = \frac{1}{2\mu_0} \int \frac{(\nabla \hat{\psi}_i)^2}{R^2} d\mathbf{r} = \frac{1}{2} \int_{S_W} \frac{1}{R^2} \hat{\psi}_i (\mathbf{n} \cdot \nabla \hat{\psi}_i) dS_W - \frac{1}{2} \int_{S_P} \frac{1}{R^2} \hat{\psi}_i (\mathbf{n} \cdot \nabla \hat{\psi}_i) dS_P \qquad (31)$$

Here and below $\mathbf{n}$ always refers to an outward pointing normal. This relation allows us to write Eq. (28) as

$$BT = \delta W_{V_i} - \frac{1}{2} \int_{S_W} \frac{1}{R^2} \hat{\psi}_i (\mathbf{n} \cdot \nabla \hat{\psi}_i) dS_W \qquad (32)$$

- **The outer vacuum region**

The boundary term can be further simplified by introducing the magnetic energy in the outer vacuum region. In analogy with Eqs. (29) - (31) we can write the outer vacuum energy (with subscript $o$) as

$$\delta W_{V_o} = \frac{1}{2\mu_0} \int (\hat{\mathbf{B}}_o)^2 \, d\mathbf{r} = \frac{1}{2\mu_0} \int \frac{(\nabla \hat{\psi}_o)^2}{R^2} \, d\mathbf{r}$$

$$= \frac{1}{2\mu_0} \int_{S_\infty} \frac{1}{R^2} \hat{\psi}_o (\mathbf{n} \cdot \nabla \hat{\psi}_o) \, dS_\infty - \frac{1}{2\mu_0} \int_{S_W} \frac{1}{R^2} \hat{\psi}_o (\mathbf{n} \cdot \nabla \hat{\psi}_o) \, dS_W \tag{33}$$

The contribution at the surface at infinity, $S_\infty$, vanishes because of the regularity boundary condition far from the plasma. Thus, Eq. (33) reduces to

$$\delta W_{V_o} + \frac{1}{2\mu_0} \int_{S_W} \frac{1}{R^2} \hat{\psi}_o (\mathbf{n} \cdot \nabla \hat{\psi}_o) \, dS_W = 0 \tag{34}$$

This expression is now added to Eq. (32) leading to

$$BT = \delta W_{V_I} + \delta W_{V_o} + \frac{1}{2\mu_0} \int_{S_W} \frac{1}{R^2} [\hat{\psi}_o (\mathbf{n} \cdot \nabla \hat{\psi}_o) - \hat{\psi}_i (\mathbf{n} \cdot \nabla \hat{\psi}_i)] \, dS_W \tag{35}$$

In Eq. (35) it is understood that the $\hat{\psi}_i$ terms are evaluated on the inner surface of the wall while the $\hat{\psi}_o$ terms are evaluated on the outer surface.

Equation (35) is in a convenient form since in the limit of a thin wall the surface terms are ultimately transformed into a simple set of jump conditions arising from the solution for the fields within the resistive wall.

• **The fields within the resistive wall**

The perturbed fields within the resistive wall are determined as follows. We write

$$\mathbf{B}_w = \nabla \times \mathbf{A}_w$$

$$\mathbf{E}_w = -\gamma \mathbf{A}_w - \nabla V_w = -\gamma \mathbf{A}_w \tag{36}$$

$$\mathbf{J}_w = \sigma \mathbf{E}_w$$

Here, $\sigma$ is the wall conductivity, the scalar electric potential $V_w$ has been set to zero as the gauge condition, and all quantities are assume to vary as $Q(\mathbf{r},t) = Q(\mathbf{r}) \exp(\gamma t)$. Ampere's Law then can be written as

$$\nabla \times \nabla \times \mathbf{A}_w = -\mu_0 \sigma \gamma \mathbf{A}_w \tag{37}$$

Now, form the dot product of Eq. (37) with $\mathbf{e}_\phi / R$ and define $\mathbf{A}_w \cdot \mathbf{e}_\phi = \psi_w / R$. A short calculation for $n = 0$ symmetry yields

$$\nabla^2 \psi_w - \frac{2}{R} \nabla \psi_w \cdot \nabla R = \mu_0 \sigma \gamma \psi_w \qquad (38)$$

It is at this point that the thin wall approximation is introduced in order to obtain an analytic solution for $\psi_w$. Assume that the wall thickness is denoted by $d$ while the minor radius of the wall at $Z = 0$ is denoted by $b \sim L_W / 2\pi$ where $L_W$ is the wall circumference. The thin wall approximation assumes that $d / b \ll 1$. The thin wall ordering is then given by

$$\mu_0 \gamma \sigma b d \sim \gamma \tau_w \sim 1$$
$$\mathbf{n} \cdot \nabla \sim \frac{1}{d} \qquad (39)$$
$$\mathbf{t} \cdot \nabla \sim \frac{1}{b}$$

The physical interpretation is as follows. The growth time $\gamma^{-1} \sim \mu_0 \sigma b d \sim \tau_w$ is the characteristic diffusion time of magnetic flux through a wall of thickness $d$ and conductivity $\sigma$ into a vacuum region of thickness $b$. The unit vector $\mathbf{n}$ is the outward normal to the wall and the ordering $\mathbf{n} \cdot \nabla \sim 1 / d$ implies rapid variation across the wall. Similarly the unit vector $\mathbf{t}$ is tangential to the wall in the poloidal direction. The ordering $\mathbf{t} \cdot \nabla \sim 1 / b$ implies that tangential variation is slower (on the scale of the device size) than normal variation.

The analytic solution to Eq. (38) can now be obtained using the resistive wall analog of the "constant $\psi$" approximation that arises in tearing mode theory [Furth et al. 1963]. We define a tangential poloidal arc length $l$ and a normal distance $s$ measured with respect to the inner surface of the resistive wall such that $0 < s < d$. This implies that

$$\mathbf{n} \cdot \nabla = \frac{\partial}{\partial s}$$
$$\mathbf{t} \cdot \nabla = \frac{\partial}{\partial l} \qquad (40)$$

The thin wall expansion for $\psi_w$ is then written as

$$\psi_w(s,l) = \overline{\psi}(l) + \tilde{\psi}(s,l) + \ldots \tag{41}$$

where $\tilde{\psi} / \overline{\psi} \sim d / b$. After some analysis we find that the first non-vanishing contribution to Eq. (38) is given by

$$\frac{\partial^2 \tilde{\psi}}{\partial s^2} = \mu_0 \sigma \gamma \overline{\psi} \tag{42}$$

This equation can be easily integrated leading to the following analytic solution for $\psi_w$

$$\psi_w(s,l) = \left(1 + \mu_0 \sigma \gamma \frac{s^2}{2} + c_1 + c_2 \frac{s}{d}\right)\overline{\psi} \tag{43}$$

where $c_1(l)$, $c_2(l)$ are two free, order $d / b$ integration constants arising from the solution to Eq. (42).

The critical information to extract from these solutions is the change in $\psi_w$ and its normal derivative across the resistive wall. These values are given by

$$\psi_w(d,l) - \psi_w(0,l) = \overline{\psi}\left(\frac{\alpha d}{2} + c_2\right) \sim \frac{d}{b}\overline{\psi} \approx 0$$

$$\tag{44}$$

$$\mathbf{n} \cdot \nabla \psi_w(d,l) - \mathbf{n} \cdot \nabla \psi_w(0,l) = \frac{\partial \psi_w(d,l)}{\partial s} - \frac{\partial \psi_w(0,l)}{\partial s} = \alpha \overline{\psi}$$

Here, $\alpha = \gamma \mu_0 \sigma d \sim 1 / b$ is related to the resistive wall growth time. Observe that there is a small, negligible jump in the flux across the wall. There is, however, a finite jump in the normal derivative, representing the current flowing in the wall.

The information in Eq. (44) is related to the inner and outer vacuum solutions by recognizing that even though the wall is thin, it still has a small finite thickness. The implication of finite thickness is that there are no ideal infinitesimally thin surface currents on either face of the wall. Therefore, across each face of the wall the flux and its normal derivative must be continuous with the adjacent vacuum fields. Specifically, coupling from the wall to the vacuum regions requires that

|          Inner Surface          |          Outer surface          |
| :---: | :---: |

$$\left[\hat{\psi}_i(s,l) - \psi_w(s,l)\right]_{s=0} = 0 \qquad\qquad \left[\hat{\psi}_o(s,l) - \psi_w(s,l)\right]_{s=d} = 0 \qquad (45)$$

$$\left[\mathbf{n}\cdot\nabla\hat{\psi}_i(s,l) - \mathbf{n}\cdot\nabla\psi_w(s,l)\right]_{s=0} = 0 \qquad \left[\mathbf{n}\cdot\nabla\hat{\psi}_o(s,l) - \mathbf{n}\cdot\nabla\psi_w(s,l)\right]_{s=d} = 0$$

By combining Eqs. (44) and (45) we obtain a set of jump conditions on the vacuum magnetic fields that takes into account the effects of the resistive wall. Simple substitution yields

$$\hat{\psi}_o(d,l) - \hat{\psi}_i(0,l) = 0$$
$$\mathbf{n}\cdot\nabla\hat{\psi}_o(d,l) - \mathbf{n}\cdot\nabla\hat{\psi}_i(0,l) = \alpha\hat{\psi}_i(0,l) \qquad (46)$$

This is the information required to complete the variational principle.

- **The final variational principle**

Return now to the expression for the boundary term given by Eq. (35). By substituting the top relation in Eq. (46) we obtain an expression for the boundary term in the thin wall limit that can be written as

$$BT = \delta W_{V_I} + \delta W_{V_O} + \frac{1}{2\mu_0}\int_{S_W}\frac{\hat{\psi}_i}{R^2}[(\mathbf{n}\cdot\nabla\hat{\psi}_o) - (\mathbf{n}\cdot\nabla\hat{\psi}_i)]dS_W \qquad (47)$$

The second natural boundary condition is introduced into Eq. (47) by substituting the bottom relation from Eq. (46),

$$BT = \delta W_{V_I} + \delta W_{V_O} + \frac{\alpha}{2\mu_0}\int_{S_W}\frac{\hat{\psi}_i^2}{R^2}dS_W \qquad (48)$$

Observe that the last term is now in a self-adjoint form.

By combining Eq. (48) with Eq. (19) we finally obtain the desired variational principle

$$\delta W = \delta W_F + \delta W_{V_I} + \delta W_{V_O} + \alpha W_D = 0$$

$$\delta W_F = \frac{1}{2\mu_0} \int_{V_P} \left[ \frac{(\nabla \psi)^2}{R^2} - \left( \mu_0 p'' + \frac{1}{2R^2} F^{2''} \right) \psi^2 \right] d\mathbf{r} + \frac{1}{2\mu_0} \int_{S_P} \left( \frac{\mu_0 J_\phi}{R^2 B_p} \psi^2 \right) dS_P$$

$$\delta W_{V_I} = \frac{1}{2\mu_0} \int_{V_I} \frac{(\nabla \hat{\psi}_i)^2}{R^2} d\mathbf{r}$$

$$\delta W_{V_O} = \frac{1}{2\mu_0} \int_{V_O} \frac{(\nabla \hat{\psi}_o)^2}{R^2} d\mathbf{r}$$

$$W_D = \frac{1}{2\mu_0} \int_{S_W} \frac{\hat{\psi}_i^2}{R^2} dS_W$$

(49)

The variational principle is very similar in form to ideal MHD. When substituting trial functions all that is necessary is to insure that the perturbed fluxes are continuous across the plasma-vacuum interface and across the resistive wall. The normal derivative requirements are automatically accounted for by means of the natural boundary conditions.

## 4. Summary

We have presented a general formulation of the $n = 0$ resistive wall stability problem. The end result is a variational principle which, as shown in the accompanying paper, is quite amenable to numerical analysis. The key new feature introduced in the analysis is the presence of a feedback system. A simple but reliable measure of the effect of feedback is determined by calculating the maximum stable $k$ and corresponding $d$ for fixed $\gamma \tau_w \propto \alpha b$ [1]. Numerical results obtained with this formalism and implications for reactor designs are given in the accompanying paper.

## Acknowledgments

The authors would like to thank Prof. Dennis Whyte (MIT) for providing the motivation for this work and for many insightful conversations. J.P. Lee and A.J. Cerfon were supported by the U.S. Department of Energy, Office of Science, Fusion Energy Sciences under Award Numbers. DE-FG02-86ER53223 and DE-SC0012398. J.

---

[1] The precise definition of the wall diffusion time, $\tau_w = \mu_0 \sigma d L_W / 2\pi$ with $L_W$ the wall circumference, will appear naturally during the discussion of the numerical analysis.

P. Freidberg was partially supported by the U.S. Department of Energy, Office of Science, Fusion Energy Sciences under Award Number DE-FG02-91ER54109.

**Appendix A**
**Relation between MHD stability and neighboring equilibria**

## 1. Introduction

We present a derivation of the fluid contribution ($\delta W_F$) to the total potential energy ($\delta W$) describing the $n = 0$ stability of a tokamak. The derivation is presented for arbitrary profiles and then simplified at the end for the Solov'ev profiles. The derivation requires a substantial amount of analysis. However, the final result is quite simple and is shown to be directly related to neighboring equilibria of the Grad-Shafranov equation.

## 2. The starting point

The starting point for the analysis is the expression for the total MHD potential energy $\delta W$ written in the "standard" form (see Eq. 19) [21]

$$\delta W(\boldsymbol{\xi}, \boldsymbol{\xi}) = \delta W_F(\boldsymbol{\xi}, \boldsymbol{\xi}) + BT$$

$$\delta W_F(\boldsymbol{\xi}, \boldsymbol{\xi}) = \frac{1}{2} \int [(\mathbf{Q})^2 + \gamma p (\nabla \cdot \boldsymbol{\xi})^2 - \boldsymbol{\xi}_\perp \cdot \mathbf{J} \times \mathbf{Q} + (\boldsymbol{\xi}_\perp \cdot \nabla p)(\nabla \cdot \boldsymbol{\xi}_\perp)] d\mathbf{r} \qquad (A.1)$$

$$BT = \frac{1}{2} \int_{S_P} (\mathbf{n} \cdot \boldsymbol{\xi}_\perp)[\mathbf{B} \cdot \mathbf{Q} - \gamma p \nabla \cdot \boldsymbol{\xi} - \boldsymbol{\xi}_\perp \cdot \nabla p] dS_P$$

Here $\mathbf{Q} = \nabla \times (\boldsymbol{\xi}_\perp \times \mathbf{B})$ and for convenience $\mu_0$ has been set to unity. It can be re-inserted at the end of the calculation. The quantity $\delta W_F$ is the fluid contribution while $BT$ is the boundary term that will ultimately be related to the vacuum energy and the resistive dissipated wall power. Note also that for the $n = 0$ mode in an up-down symmetric tokamak we can assume that $\boldsymbol{\xi}$ is purely real.

The present analysis is focused solely on obtaining a simple form for $\delta W_F$. Observe that even though our goal is to calculate the eigenvalue $\omega$ for resistive wall modes, we can neglect the $\omega^2$ contribution due to the plasma inertia. The reason is that resistive wall growth rates are much slower than ideal MHD growth rates, thus justifying the neglect of MHD inertial effects. This is the explanation of why only $\delta W(\boldsymbol{\xi}, \boldsymbol{\xi})$ and not $\omega^2 K(\boldsymbol{\xi}, \boldsymbol{\xi})$ is needed to describe the plasma behavior. The eigenvalue $\omega$ will appear in the evaluation of $BT$.

## 3. Incompressibility

The first step in the analysis is to examine the plasma compressibility term. As is well known this is the only term in which $\xi_\parallel$ appears. The question is whether or not a well behaved $\xi_\parallel$ can be found that makes $\nabla \cdot \boldsymbol{\xi} = 0$ which, if possible, clearly minimizes the plasma compressibility term.

To answer this question we have to make use of the following symmetries with respect to the $Z$ dependence of the equilibrium magnetic fields

$$B_R(R, -Z) = -B_R(R, Z)$$
$$B_Z(R, -Z) = +B_Z(R, Z) \qquad \text{(A.2)}$$
$$B_\phi(R, -Z) = +B_\phi(R, Z)$$

Now, for the $n = 0$ mode a general perturbation $\boldsymbol{\xi}$ can be always be written as the sum of an "even $\xi_Z$" contribution plus an "odd $\xi_Z$" contribution which are completely decoupled from one another because of the equilibrium symmetry. The perturbation symmetries are as follows,

Even $\xi_Z$ Symmetry  $\qquad\qquad$ Odd $\xi_Z$ Symmetry

$$\xi_Z(R, -Z) = +\xi_Z(R, Z) \qquad\qquad \xi_Z(R, -Z) = -\xi_Z(R, Z)$$
$$\xi_R(R, -Z) = -\xi_R(R, Z) \qquad\qquad \xi_R(R, -Z) = +\xi_R(R, Z) \qquad \text{(A.3)}$$
$$\xi_\phi(R, -Z) = +\xi_\phi(R, Z) \qquad\qquad \xi_\phi(R, -Z) = -\xi_\phi(R, Z)$$
$$\xi_\parallel(R, -Z) = +\xi_\parallel(R, Z) \qquad\qquad \xi_\parallel(R, -Z) = -\xi_\parallel(R, Z)$$
$$\nabla \cdot \boldsymbol{\xi}_\perp(R, -Z) = -\nabla \cdot \boldsymbol{\xi}_\perp(R, -Z) \qquad\qquad \nabla \cdot \boldsymbol{\xi}_\perp(R, -Z) = +\nabla \cdot \boldsymbol{\xi}_\perp(R, -Z)$$

The $n = 0$ modes of interest have even $\xi_Z$ symmetry (e.g. $x_Z = \text{constant}$) corresponding to "vertical displacements". The less interesting modes have odd $\xi_Z$ symmetry (e.g. $\xi_R = \text{constant}$) and represent "horizontal displacements".

For the vertical displacements of interest the most unstable modes are always incompressible. To show this we note that for a minimizing incompressible displacement $\xi_\parallel$ must satisfy

$$\mathbf{B} \cdot \nabla \frac{\xi_\parallel}{B} = \mathbf{B}_p \cdot \nabla \frac{\xi_\parallel}{B} = B_p \frac{\partial}{\partial l} \frac{\xi_\parallel}{B} = -\nabla \cdot \boldsymbol{\xi}_\perp \qquad (A.4)$$

where $\mathbf{B}_p$ is the poloidal magnetic field and $l$ is poloidal arc length. Thus, for $\xi_\parallel$ to be well behaved (i.e. be periodic) it must satisfy the periodicity constraint

$$0 = \oint \frac{\partial}{\partial l} \left( \frac{\xi_\parallel}{B} \right) dl = -\oint \frac{\nabla \cdot \boldsymbol{\xi}_\perp}{B_p} dl = -\oint \frac{\nabla \cdot \boldsymbol{\xi}_\perp}{B_Z} dZ \qquad (A.5)$$

This constraint is automatically satisfied for the even $\xi_Z$ symmetry displacement.

The conclusion is that the plasma compressibility term in $\delta W_F$ is minimized by choosing $\xi_\parallel$ to satisfy Eq. (A.4) leading to the result $\gamma p (\nabla \cdot \boldsymbol{\xi})^2 = 0$. The remainder of $\delta W_F$ is only a function of $\boldsymbol{\xi}_\perp$.

### 4. Reformulating $\delta W_F$ in terms of the vector potential

The evaluation of $\delta W_F$ is more conveniently carried out in terms of the perturbed vector potential $\mathbf{A}_{\perp 1}$ rather than the plasma displacement $\boldsymbol{\xi}_\perp$. The familiar relation between these two quantities is

$$\mathbf{A}_{\perp 1} = \boldsymbol{\xi}_\perp \times \mathbf{B} \qquad \boldsymbol{\xi}_\perp = \frac{\mathbf{B} \times \mathbf{A}_{\perp 1}}{B^2} \qquad (A.6)$$

In general $\mathbf{A}_{\perp 1}$ can be vector decomposed as follows

$$\mathbf{A}_{\perp 1} = f(R,Z) \nabla \Psi + g(R,Z) \mathbf{B}_p + h(R,Z) \mathbf{e}_\phi \qquad (A.7)$$

Here, $\Psi(R,Z)$ is the equilibrium flux function satisfying the Grad-Shafranov equation. We next make use of the constraint $\mathbf{A}_{\perp 1} \cdot \mathbf{B} = 0$ and introduce the perturbed flux $\psi(R,Z) = RA_{\phi 1}(R,Z) = Rh(R,Z)$. A short calculation then enables us to rewrite $\mathbf{A}_{\perp 1}$ as

$$\mathbf{A}_{\perp 1} = f \nabla \Psi - \frac{F\psi}{R^2 B_p^2} \mathbf{B}_p + \frac{\psi}{R} \mathbf{e}_\phi \qquad (A.8)$$

where $F(\psi) = RB_\phi$. The basic unknowns in the problem are $\psi, f$.

Before proceeding with the evaluation of $\delta W_F$ it is useful to list a number of relations involving $\mathbf{A}_{\perp 1}$ that enter the analysis. These are relatively straightforward to derive.

$$\mathbf{Q} = \nabla \times \mathbf{A}_{\perp 1} = \frac{1}{R} \nabla \psi \times \mathbf{e}_\phi + \nabla f \times \nabla \Psi + \left[ \frac{1}{R} \nabla \Psi \cdot \nabla \left( \frac{F\psi}{R^2 B_p^2} \right) - \frac{FJ_\phi \psi}{R^2 B_p^2} \right] \mathbf{e}_\phi$$

$$\mathbf{Q} \cdot \mathbf{A}_{\perp 1} = \frac{\psi}{R} \left[ \frac{1}{R} \nabla \Psi \cdot \nabla \left( \frac{F\psi}{R^2 B_p^2} \right) - \frac{FJ_\phi \psi}{R^2 B_p^2} \right] - \frac{F\psi}{R^4 B_p^2} \nabla \Psi \cdot \nabla \psi$$

$$+ 2\psi (\mathbf{B}_p \cdot \nabla f) - \nabla \cdot (\psi f \, \mathbf{B}_p)$$

$$\mathbf{Q} \cdot \mathbf{B} = \frac{1}{R^2} \nabla \Psi \cdot \nabla \psi + \frac{F}{R} \left[ \frac{1}{R} \nabla \Psi \cdot \nabla \left( \frac{F\psi}{R^2 B_p^2} \right) - \frac{FJ_\phi \psi}{R^2 B_p^2} \right] + F(\mathbf{B}_p \cdot \nabla f) \qquad \text{(A.9)}$$

$$\nabla \psi \cdot (\mathbf{e}_\phi \times \mathbf{A}_{\perp 1}) = -\frac{F\psi}{R^3 B_p^2} \nabla \Psi \cdot \nabla \psi - Rf \, \mathbf{B}_p \cdot \nabla \psi$$

$$\boldsymbol{\xi}_\perp \cdot \nabla p = -p' \psi$$

$$\mathbf{A}_{\perp 1} \cdot (\nabla \Psi \times \mathbf{B}) = -B^2 \psi$$

with prime denoting $d / d\Psi$ as in the main text. We are now ready to evaluate $\delta W_F$

## 5. Evaluation of $\delta W_F$

There are three terms appearing in $\delta W_F$ as given by Eq. (A.1) after setting $\nabla \cdot \boldsymbol{\xi} = 0$. Using Eq. (A.9) these terms can be evaluated in a straightforward way. The magnetic energy term is given by

$$(\mathbf{Q})^2 = \frac{(\nabla \psi)^2}{R^2} + \left[ \frac{1}{R} \nabla \Psi \cdot \nabla \left( \frac{F\psi}{R^2 B_p^2} \right) - \frac{FJ_\phi \psi}{R^2 B_p^2} \right]^2$$

$$+ (R\mathbf{B}_p \cdot \nabla f)^2 + 2(R\mathbf{B}_p \cdot \nabla f) \left[ \frac{1}{R} \nabla \Psi \cdot \nabla \left( \frac{F\psi}{R^2 B_p^2} \right) - \frac{FJ_\phi \psi}{R^2 B_p^2} \right] \qquad \text{(A.10)}$$

The current term is evaluated by making use of the equilibrium relation $\mathbf{J} = F'\mathbf{B} + Rp'\mathbf{e}_\phi$ which leads to a fairly complicated expression

$$
\begin{aligned}
-\boldsymbol{\xi}_\perp \cdot (\mathbf{J} \times \mathbf{Q}) = & -\left(F' + F\frac{p'}{B^2}\right)(\mathbf{A}_{\perp 1} \cdot \mathbf{Q}) + \frac{p'}{B^2}\psi(\mathbf{B} \cdot \mathbf{Q}) \\
= & -\left(F' + F\frac{p'}{B^2}\right)\left\{\frac{\psi}{R}\left[\frac{1}{R}\nabla\Psi \cdot \nabla\left(\frac{F\psi}{R^2 B_p^2}\right) - \frac{FJ_\phi\psi}{R^2 B_p^2}\right] - \frac{F\psi}{R^4 B_p^2}\nabla\Psi \cdot \nabla\psi \right. \\
& \left. + 2\frac{\psi}{R}(R\mathbf{B}_p \cdot \nabla f) - \nabla \cdot (\psi f\,\mathbf{B}_p)\right\} \\
& + \frac{p'\psi}{B^2}\left\{\frac{1}{R^2}\nabla\Psi \cdot \nabla\psi + \frac{F}{R}\left[\frac{1}{R}\nabla\Psi \cdot \nabla\left(\frac{F\psi}{R^2 B_p^2}\right) - \frac{FJ_\phi\psi}{R^2 B_p^2}\right] + \frac{F}{R}(R\mathbf{B}_p \cdot \nabla f)\right\}
\end{aligned}
\tag{A.11}
$$

The remaining pressure term requires some minor algebraic manipulations. The result is

$$
\begin{aligned}
(\boldsymbol{\xi}_\perp \cdot \nabla p)(\nabla \cdot \boldsymbol{\xi}_\perp) = & \nabla \cdot [(\boldsymbol{\xi}_\perp \cdot \nabla p)\boldsymbol{\xi}_\perp] - \boldsymbol{\xi}_\perp \cdot \nabla(\boldsymbol{\xi}_\perp \cdot \nabla p) \\
= & -\nabla \cdot (p'\psi\boldsymbol{\xi}_\perp) + \frac{p''\psi}{B^2}[\mathbf{A}_{\perp 1} \cdot (\nabla\Psi \times \mathbf{B})] + \frac{p'}{B^2}[\mathbf{A}_{\perp 1} \cdot (\nabla\psi \times \mathbf{B})] \\
= & -\nabla \cdot (p'\psi\boldsymbol{\xi}_\perp) - p''\psi^2 + \frac{p'F}{RB^2}[\nabla\psi \cdot (\mathbf{e}_\phi \times \mathbf{A}_{\perp 1})] - \frac{p'\psi}{R^2 B^2}(\nabla\Psi \cdot \nabla\psi) \\
= & -\nabla \cdot (p'\psi\boldsymbol{\xi}_\perp) - p''\psi^2 - \frac{p'\psi}{R^2 B_p^2}(\nabla\Psi \cdot \nabla\psi) - \frac{p'F}{RB^2}[f(R\mathbf{B}_p \cdot \nabla\psi)]
\end{aligned}
\tag{A.12}
$$

The next task is to sum these three complicated contributions and simplify the result. The terms in the sum can be collected and written as follows

$$
\begin{aligned}
\delta W_F(\boldsymbol{\xi},\boldsymbol{\xi}) = & \frac{1}{2}\int [(\mathbf{Q})^2 - \boldsymbol{\xi}_\perp \cdot \mathbf{J} \times \mathbf{Q} + (\boldsymbol{\xi}_\perp \cdot \nabla p)(\nabla \cdot \boldsymbol{\xi}_\perp)]d\mathbf{r} = \frac{1}{2}\int U\,d\mathbf{r} \\
& U = U_1(f,f) + U_2(f,\psi) + U_3(\psi,\psi)
\end{aligned}
\tag{A.13}
$$

Here, each $U_j$ is a combination of algebraic and differential operators acting on the terms appearing in the arguments. After a short calculation the first two terms on the right hand side can be simplified leading to

$$U_1(f,f) + U_2(f,\psi) = (R\mathbf{B}_p \cdot \nabla f)^2 + \nabla \cdot (F'f\psi\mathbf{B}_p)$$
$$+ 2(R\mathbf{B}_p \cdot \nabla f)\left[\frac{1}{R}\nabla\Psi \cdot \nabla\left(\frac{F\psi}{R^2 B_p^2}\right) - \frac{FJ_\phi\psi}{R^2 B_p^2} - \frac{F'\psi}{R}\right] \qquad \text{(A.14)}$$

The divergence term integrates to zero over the plasma surface. The $f$ dependence of the remaining terms involves only $(R\mathbf{B}_p \cdot \nabla f)$. These terms are simplified by completing the square on $(R\mathbf{B}_p \cdot \nabla f)$. The fluid energy $\delta W_F$ is then minimized by choosing

$$(R\mathbf{B}_p \cdot \nabla f) = -\frac{1}{R}\nabla\Psi \cdot \nabla\left(\frac{F\psi}{R^2 B_p^2}\right) + \frac{FJ_\phi\psi}{R^2 B_p^2} + \frac{F'\psi}{R} \qquad \text{(A.15)}$$

Note that as for the incompressibility analysis, the periodicity constraint on $(R\mathbf{B}_p \cdot \nabla f)$ is automatically satisfied for the perturbation with even $\xi_Z$ symmetry. The end result is that

$$U_1(f,f) + U_2(f,\psi) = -\left[\frac{1}{R}\nabla\Psi \cdot \nabla\left(\frac{F\psi}{R^2 B_p^2}\right) - \frac{FJ_\phi\psi}{R^2 B_p^2} - \frac{F'\psi}{R}\right]^2 \qquad \text{(A.16)}$$

which is only a function of $\psi$.

The complete integrand $U$ is now evaluated by adding Eq. (A.16) to $U_3(\psi,\psi)$. After another tedious calculation we obtain

$$U = \frac{(\nabla\psi)^2}{R^2} - \left(p'' + \frac{F'^2}{R^2} + \frac{FF'J_\phi}{R^3 B_p^2}\right)\psi^2 - \nabla \cdot (p'\psi\boldsymbol{\xi}_\perp)$$
$$+ \frac{F'\psi}{R^2}\left[\nabla\Psi \cdot \nabla\left(\frac{F\psi}{R^2 B_p^2}\right) + \frac{F}{R^2 B_p^2}(\nabla\Psi \cdot \nabla\psi)\right] \qquad \text{(A.17)}$$

After some further manipulations that make use of the equilibrium Grad-Shafranov equation we find that the term on the second line of Eq. (A.17) can be rewritten as

$$\frac{F'\psi}{R^2}\left[\nabla\Psi \cdot \nabla\left(\frac{F\psi}{R^2 B_p^2}\right) + \frac{F}{R^2 B_p^2}(\nabla\Psi \cdot \nabla\psi)\right] = \nabla \cdot \left(\frac{FF'\psi^2}{R^4 B_p^2}\nabla\Psi\right) - \left(\frac{FF''}{R^2} - \frac{FF'J_\phi}{R^3 B_p^2}\right)\psi^2 \qquad \text{(A.18)}$$

Equation (A.18) is substituted into Eq. (A.17). The divergence terms are converted into surface integrals and then simplified using the equilibrium Grad-Shafranov equation. This leads to the final desired form of $\delta W_F$

$$\delta W_F = \frac{1}{2} \int \left[ \frac{(\nabla \psi)^2}{R^2} - \left( p'' + \frac{1}{2R^2} F^{2''} \right) \psi^2 \right] d\mathbf{r} + \frac{1}{2} \int_{S_p} \left( \frac{J_\phi}{R^2 B_p} \psi^2 \right) dS \qquad (A.19)$$

There are three points worth noting. First, a simple variational analysis shows that the function $\psi$ that minimizes the volume contribution satisfies

$$\Delta^* \psi = -\left( R^2 p'' + \frac{1}{2} F^{2''} \right) \psi \qquad (A.20)$$

This equation is identical to the perturbed Grad-Shafranov equation corresponding to neighboring equilibria. Second, the surface contribution is nonzero only if there is a jump in the edge current density. For many profiles $J_\phi$ is zero at the plasma edge but it is finite for the Solov'ev profiles. Third, there is a great simplification in the volume contribution for the Solov'ev model in that $p'' = F^{2''} = 0$. Thus the fluid energy for the Solov'ev model reduces to

$$\delta W_F = \frac{1}{2} \int \left[ \frac{(\nabla \psi)^2}{R^2} \right] d\mathbf{r} + \frac{1}{2} \int_{S_p} \left( \frac{J_\phi}{R^2 B_p} \psi^2 \right) dS \qquad (A.21)$$

with $\psi$ satisfying

$$\Delta^* \psi = 0 \qquad (A.22)$$

Assuming a solution to this equation can be found then the volume contribution can be converted into a surface integral. The fluid energy can then be written solely in terms of a surface integral

$$\delta W_F = \frac{1}{2} \int_{S_p} \left( \frac{1}{R^2} \psi \, \mathbf{n} \cdot \nabla \psi + \frac{J_\phi}{R^2 B_p} \psi^2 \right) dS \qquad (A.23)$$

This form is actually valid for any equilibrium profile. The beauty of the Solov'ev profiles is that $\mathbf{n} \cdot \nabla \psi$ on the plasma surface can be conveniently evaluated in terms of $\psi$ on the surface using Green's theorem since we know the analytic form of the Green's function corresponding to Eq. (A.22). For general profiles this procedure is not as convenient since we do not know the Green's function in a simple analytic form.

**Figure Captions – Paper 1**

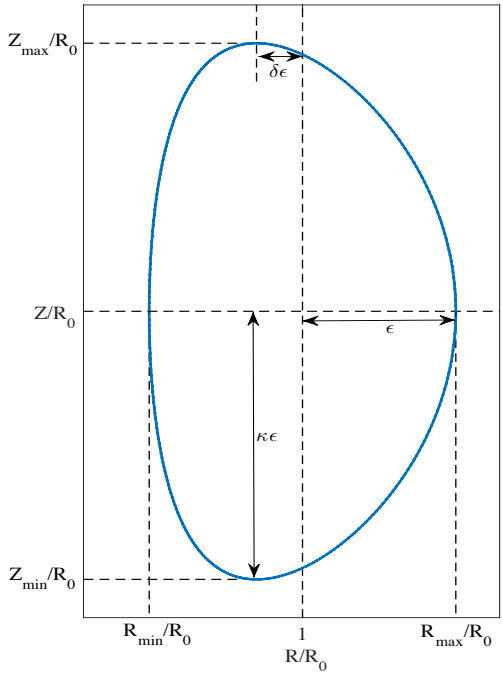

Figure 1 Geometry of the plasma equilibrium

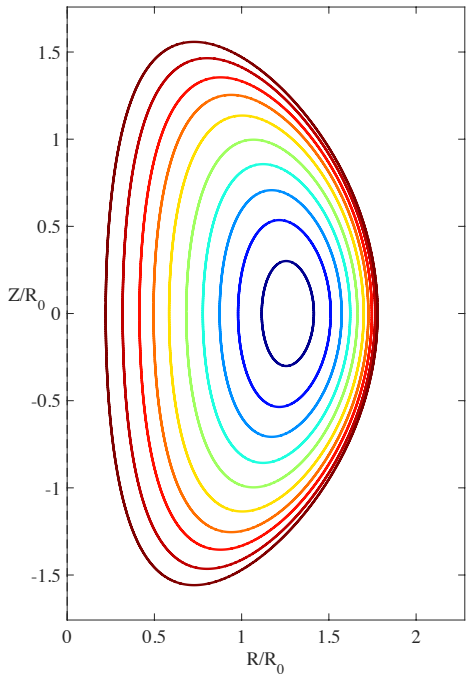

Figure 2 Flux surfaces for NSTX using the Solov'ev model. The value of $A$ has been chosen such that $J_\phi(1-\varepsilon, 0) = 0$